\documentclass[pra,twocolumn,showpacs]{revtex4}
\usepackage{latexsym}
\usepackage{epsfig}

\def\op#1{\hat{#1}}
\def\ket#1{| #1 \rangle}
\def\bra#1{\langle #1 |}
\def\ave#1{\langle #1 \rangle}
\def\vec#1{{\bf #1}}

\def\ip#1#2{\langle #1 \mid  #2 \rangle} 
\def\Tr{{\rm Tr}}

\def\ra{\rightarrow}

\begin{document}
\bibliographystyle{prsty}
\title{Quantum control using sequences of simple control pulses}
\author{S.~G.\ Schirmer}
\email{s.g.schirmer@open.ac.uk}
\author{Andrew D.\ Greentree}
\email{a.d.greentree@open.ac.uk}
\affiliation{Quantum Processes Group, 
             The Open University, 
             Milton Keynes, MK7 6AA, 
             United Kingdom}
\author{Viswanath Ramakrishna}
\email{vish@utdallas.edu}
\affiliation{Center for Engineering Math and the Program in Math Sciences, EC 35, 
             University of Texas at Dallas, 
             Richardson, Texas 75083}
\author{Herschel Rabitz}
\email{hrabitz@chemvax.princeton.edu}
\affiliation{Department of Chemistry, 
             Frick Laboratories, 
             Princeton University, 
             Princeton, NJ 08544} 
\date{May 14, 2001} 
\begin{abstract}
Structured decompositions of a desired unitary evolution operator are employed to derive
control schemes that achieve certain control objectives for finite-level quantum systems
using only sequences of simple control pulses such as square-waves with finite rise and 
decay times or Gaussian wavepackets.  The technique is applied to find control schemes 
that achieve population transfers for pure-state systems, complete inversions of the
ensemble populations for mixed-state systems, create arbitrary superposition states and
optimize the ensemble average of observables.
\end{abstract}
\pacs{03.65.Bz}
\maketitle

\section{Introduction}
\label{sec:intro}

The ability to control quantum systems is an essential prerequisite for many present 
and potential future applications involving the manipulation of atomic and molecular 
quantum states \cite{SCI288p824}.  Some of the important applications are quantum state
engineering \cite{PRA63n023408}, control of chemical reactions \cite{JCP113p3510} and 
laser cooling of internal molecular degrees of freedom \cite{FD113p365,PRA63n069101}. 
Quantum control theory may also reveal new ways to solve problems crucial to quantum 
computing such as initializing quantum registers \cite{qph0104030} or building robust
quantum memory \cite{qph0103118}.

Due to the wide range of applications, the immediate aims of quantum control may vary.  
However, the control objective can usually be classified as one of the following:
\begin{enumerate}
\item to steer the system from its initial state to a target state with desired
      properties, 
\item to maximize (or minimize) the expectation value (ensemble average) of a 
      selected observable, or
\item to achieve a certain evolution of the system.  
\end{enumerate}
Despite their apparent dissimilarity, these control objectives are closely related. 
Indeed, (1) can be considered a special case of (2) in which the observable is the 
projector onto the subspace spanned by the target state and (2) is a special case 
of (3) where we desire to find an evolution operator that maximizes the expectation 
value of the selected observable either at a specific target time or at some time in
the future.  Thus, one of the central problems of quantum control is to achieve a 
desired evolution of the system by applying external control fields and the primary
challenge is to find control pulses (or sequences of such pulses) that are feasible
from a practical point of view and effective in realizing the control objective.

Many of the control strategies for quantum systems that have been proposed to date rely
on numerical methods to solve the control or optimization problem \cite{JPA33p4643, 
PRA61n012101,PRA62n012105, PRA62n012307, JCP109p9318, JCP110p34, JCP110p9825}.  In this
paper, we explore an alternative approach to quantum control, which is based on explicit
generation of unitary operators using Lie group decompositions \cite{PRA61n032106}.  The
technique uses the rotating wave approximation (RWA) and employs frequency discrimination
or other atomic selection rules to address transitions individually.  Although these
assumptions seem to preclude the application of the technique to some model systems of 
interest, e.g., systems with equally spaced or almost equally spaced energy levels, it 
should nevertheless be applicable to many quantum systems.  A significant advantage of
the approach we pursue in this paper is the flexibility of the control pulse shapes, 
which makes it possible to choose control pulses that are reasonably simple to create in
the laboratory, such as square-wave pulses with finite rise and decay times or Gaussian 
wavepackets.

\section{Mathematical formulation of the control problem}
\label{sec:basics}

We consider a quantum system with a discrete energy spectrum whose energy levels $E_n$ 
are assumed to be finite, non-degenerate and ordered in an increasing sequence,
\begin{equation} \label{eq:En}
  E_1 < E_2 < \cdots < E_N,
\end{equation}
where $N<\infty$ is the dimension of the Hilbert space of pure states of the system.  
Note that non-degeneracy and the particular ordering of the energy levels are assumed
primarily for simplicity.  These requirements can often be relaxed if selection rules
are available to distinguish degenerate energy levels and different transitions that 
have the same transition frequency.  Moreover, non-degeneracy of the energy levels can
often be assured by considering suitable subspaces of the Hilbert space of pure states.
Our model is strictly valid only for dissipation-free single atoms or optically thin 
ensembles of homogeneously broadened atoms.  However, by choosing appropriately short 
control times (with respect to the lifetimes of the excited states) \cite{qph0103117}, 
our idealized model should be applicable to a variety of quantum systems including many
atomic and molecular $N$-state systems, Rydberg atoms or particles in (anharmonic) 
quantum wells.

\begin{figure}
\epsfbox{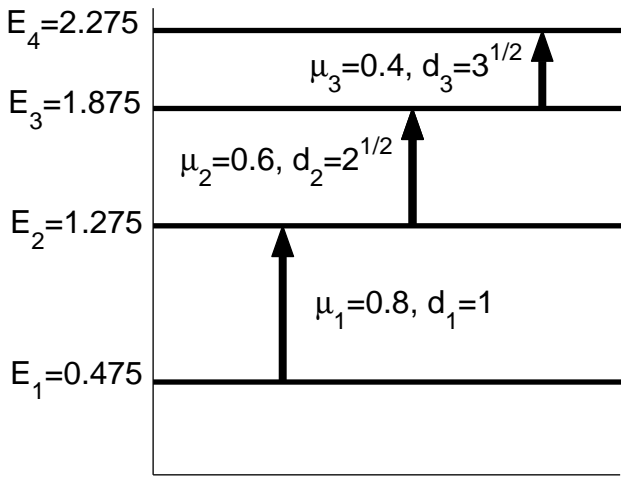}
\caption{Energy-level and transition diagram for a four-level Morse oscillator with 
anharmonicity $\alpha=0.1$.}  \label{Fig:Sys}
\end{figure}

Although the results presented in this paper do not depend on a specific model, for clarity,
we shall illustrate our results using a four-level Morse oscillator with energy levels
\begin{equation}\label{eq:EnMorse}
  E_n= \hbar\omega_0\left(n-\frac{1}{2}\right)
       \left[1-\alpha\left(n-\frac{1}{2}\right)\right]
\end{equation}
for $1\le n\le 4$ and transition dipole moments 
\begin{equation} \label{eq:dn}
  d_n=p_{12}\sqrt{n}, \quad 1\le n\le 3,
\end{equation}
where $\omega_0$ and $p_{12}$ are constants representing the oscillator frequency and the
transition dipole moment of the transition $\ket{1}\rightarrow\ket{2}$, respectively.  The
parameter $\alpha$ determines the anharmonicity of the system, which we arbitrarily set to
$0.1$.  The energy-level and transition diagram for this system is shown in figure 
\ref{Fig:Sys}.  

The free evolution of the system is governed by the internal Hamiltonian, whose spectral
representation is
\begin{equation} \label{eq:Hzero}
  \op{H}_0 = \sum_{n=1}^N E_n \ket{n}\bra{n},
\end{equation}
where $\{\ket{n}: n=1,\ldots,N\}$ is a complete set of orthonormal eigenstates that
satisfy the stationary Schr{\"o}dinger equation
\begin{equation} \label{eq:SSE}
   \op{H}_0 \ket{n} = E_n \ket{n}, \quad 1\le n\le N.
\end{equation}

The application of external control fields perturbs the system and gives rise to a new
Hamiltonian $\op{H}=\op{H}_0+\op{H}_I$, where $\op{H}_I$ is an interaction term.  For a
control field of the form
\begin{eqnarray} \label{eq:fm}
  f_m(t) &=& 2 A_m(t) \cos(\mu_m t+\phi_m) \nonumber \\
         &=& A_m(t) \left[e^{i(\mu_m t+\phi_m)} + e^{-i(\mu_m t+\phi_m)}\right],
\end{eqnarray}
which is resonant with the transition frequency $\mu_m=E_{m+1}-E_m$ and drives only the
transition $\ket{m}\rightarrow \ket{m+1}$, using the rotating wave approximation (RWA)
leads to the following interaction term
\begin{eqnarray}
  \op{H}_m(f_m) 
  &=&   A_m(t) e^{ i(\mu_m t + \phi_m)} d_m \ket{m}\bra{m+1} \nonumber \\
  & & + A_m(t) e^{-i(\mu_m t + \phi_m)} d_m \ket{m+1}\bra{m} \label{eq:Hm},
\end{eqnarray}
where $d_m$ is the transition dipole moment for the transition $\ket{m} \rightarrow 
\ket{m+1}$, $\phi_m$ is the initial phase and $2 A_m(t)$ is the envelope of the field, 
which must be slowly varying with respect to  $\mu_m^{-1}$.  It should be noted that 
the validity of this approximation for the interaction with the field $f_m$ depends on
one's ability to individually address the $m$th transition, either by use of selection 
rules or frequency discrimination.

For frequency-selective control the transition frequencies $\mu_m$ must be sufficiently
distinct to assure that each field is resonant only with a single transition and that
off-resonant effects on other transitions are small.  To satisfy this requirement, we 
note that if a monochromatic field of constant amplitude $2A_m$ and frequency $\mu_m$ 
is applied then off-resonant effects can generally be neglected if the Rabi frequency 
\begin{equation}
  \Omega_m \equiv 2 A_m d_m
\end{equation}
is considerably less than the detuning of the field from off-resonant transitions.  
Although the situation is somewhat more complicated for fields with varying amplitude,
off-resonant effects should remain small as long as 
\begin{equation} \label{eq:detuning1}
  \max_{t} \left[2 A_m(t) d_m \right] \ll \mu_m-\mu_n, \quad \forall n \neq m,
\end{equation}
i.e., as long as the peak amplitude $2 A_m$ of the control field $f_m$ is sufficiently
small.  Since the detuning of adjacent transitions for a Morse oscillator with 
anharmonicity $\alpha$ is $\mu_m-\mu_{m+1}=2\alpha$, the peak amplitudes of control
pulses should be chosen such that
\begin{equation} \label{eq:detuning2}
   A_m \ll \alpha/d_m.
\end{equation}

The evolution of the controlled system is determined by a time-evolution operator 
$\op{U}(t)$, which satisfies the Schr{\"o}dinger equation
\begin{equation} \label{eq:SE1}
  i\frac{d}{dt}\op{U}(t) =
  \left\{ \op{H}_0 + \sum_{m=1}^M \op{H}_m[f_m(t)] \right\} \op{U}(t),
\end{equation}
as well as the initial condition $\op{U}(0)=\op{I}$, where $\op{I}$ is the identity
operator.  Note that throughout this paper we shall choose units such that $\hbar=1$.
Our aim is to achieve a certain evolution of the system by applying a sequence of 
simple control pulses.  Concretely, we seek to dynamically realize a desired unitary 
operator $\op{U}$ at a certain target time $T$, i.e.,
\begin{equation} \label{eq:UT}
  \op{U} = \op{U}(T).
\end{equation}
In some cases, we may not wish to specify a target time in advance, in which case
we simply attempt to achieve the control objective at some later time $T>0$.

\section{Determination of pulse sequence using Lie group decompositions}
\label{sec:Lie}

To solve the problem of finding the right sequence of control pulses, we apply the
interaction picture decomposition of the time-evolution operator $\op{U}(t)$,
\begin{equation} \label{eq:IPD}
  \op{U}(t) = \op{U}_0(t)\op{U}_I(t),
\end{equation}
where $\op{U}_0(t)$ is the time-evolution operator of the unperturbed system,
\begin{equation} \label{eq:U0}
  \op{U}_0(t) = \exp\left(-i\op{H}_0 t\right)
                = \sum_{n=1}^N e^{-i E_n t} \ket{n}\bra{n}
\end{equation}
and $\op{U}_I(t)$ comprises the interaction with the control fields.  Inserting
(\ref{eq:IPD}) into the Schr{\"o}dinger equation (\ref{eq:SE1}) gives
\begin{eqnarray*}
{i\frac{d}{dt}\op{U}(t)} 
 &=&      {\op{H}_0\op{U}_0(t)\op{U}_I(t)+\op{U}_0(t) i\frac{d}{dt}\op{U}_I(t)} \\
 &\doteq& {\op{H}_0\op{U}(t)+\sum_{m=1}^M \op{H}_m[f_m(t)] \op{U}_0(t)\op{U}_I(t),}
\end{eqnarray*}
which leads to 
\begin{equation} \label{eq:SE2}
  i\frac{d}{dt} \op{U}_I(t) 
 = \op{U}_0(t)^\dagger \left\{\sum_{m=1}^M \op{H}_m[f_m(t)] \right\} \op{U}_0(t) 
    \op{U}_I(t).
\end{equation}
Inserting equations (\ref{eq:U0}) and (\ref{eq:Hm}) into the right hand side of equation
(\ref{eq:SE2}) leads after some simplification (see appendix \ref{appendix:A}) to
\begin{equation}\label{eq:Omega}
  \frac{d}{dt}\op{U}_I(t)
 = \sum_{m=1}^M A_m(t) d_m \left(\sin\phi_m \op{x}_m - \cos\phi_m \op{y}_m \right) 
   \op{U}_I(t)
\end{equation}
where we define
\begin{equation}
\begin{array}{rcl}
  \op{e}_{m,n} &\equiv& \ket{m}\bra{n} \\
  \op{x}_m     &\equiv&   \op{e}_{m,m+1} - \op{e}_{m+1,m} \\
  \op{y}_m     &\equiv& i(\op{e}_{m,m+1} + \op{e}_{m+1,m}).
\end{array}
\end{equation}
Hence, if we apply a control pulse
\begin{eqnarray}
   f_k(t) &=& 2 A_k(t) \cos(\mu_m t+\phi_k) \nonumber\\
          &=& A_k(t) \left[ e^{i(\mu_m t+\phi_k)} + e^{-i(\mu_m t+\phi_k)} \right],
\end{eqnarray}
which is resonant with the transition frequency $\mu_m$, for a time period $t_{k-1}\le 
t\le t_k$ and no other control fields are applied during this time period then we have
\begin{equation}
  \op{U}_I(t)=\op{V}_k(t)\op{U}_I(t_{k-1}),
\end{equation}
where the operator $\op{V}_k(t)$ is
\begin{equation} \label{eq:Vk}
  \op{V}_k(t)
  = \exp\left[\int_{t_{k-1}}^t \!\!\!\!\!\! A_k(\tau) \, d\tau \; d_m 
    (\sin\phi_k\op{x}_m-\cos\phi_k\op{y}_m)\right].
\end{equation}
Thus, if we partition the time interval $[0,T]$ into $K$ subintervals $[t_{k-1},t_k]$
such that $t_0=0$ and $t_K=T$ and apply a sequence of fixed-frequency control pulses,
each resonant with one transition frequency $\mu_m=\mu_{\sigma(k)}$, such that during
each time interval only one control field is applied, then we have
\begin{equation}
  \op{U}(T) = \op{U}_0(T)\op{U}_I(T)
            = e^{-i\op{H}_0 T}\op{V}_K \op{V}_{K-1} \cdots \op{V}_1,
\end{equation}
where the factors $\op{V}_k$ are
\begin{equation} \label{eq:Vk1}
  \op{V}_k = \exp\left[\int_{t_{k-1}}^{t_k}\!\!\!\!\! A_k(\tau)\,d\tau\; d_{\sigma(k)}
             (\sin\phi_k\op{x}_{\sigma(k)}-\cos\phi_k\op{y}_{\sigma(k)})\right],
\end{equation}
$2 A_k(t)$ is the envelope of the $k$th pulse and $\sigma$ is a mapping from the index
set $\{1,\ldots,K\}$ to the control index set $\{1,\ldots,M\}$ that determines which of 
the control fields is active for $t\in [t_{k-1},t_k]$.

It has been shown in \cite{PRA61n032106} that any unitary operator $\op{U}$ can be 
decomposed into a product of operators of the type $\op{V}_k$ and a phase factor
$e^{i\Gamma}=\det\op{U}$, i.e., there exists a positive real number $\Gamma$ and real
numbers $C_k$ and $\phi_k$ for $1\le k\le K$ and a mapping $\sigma$ from the index set
$\{1,\ldots,K\}$ to the control-sources index set $\{1,\ldots,M\}$ such that 
\begin{equation} \label{eq:Udecomp}
  \op{U}= e^{i\Gamma}\op{V}_K\op{V}_{K-1}\cdots \op{V}_1,
\end{equation}
where the factors are
\begin{equation} \label{eq:Vk2}
  \op{V}_k = 
  \exp\left[C_k (\sin\phi_k\op{x}_{\sigma(k)}-\cos\phi_k\op{y}_{\sigma(k)})\right].
\end{equation}
This decomposition of the target operator into a product of generators of the dynamical
Lie group determines the sequence in which the lasers are to be turned on and off.  A 
general algorithm to determine the Lie group decomposition for an arbitrary operator 
$\op{U}$ is described in appendix \ref{appendix:Udecomp}.

Note that in many cases the target operator $\op{U}$ is unique only up to phase factors,
i.e., two unitary operators $\op{U}_1$ and $\op{U}_2$ in $U(N)$ are equivalent if there 
exist values $\theta_n\in [0,2\pi]$ for $1\le n\le N$ such that
\begin{equation}
  \op{U}_2 = \op{U}_1 \left(\sum_{n=1}^N e^{i\theta_n} \ket{n}\bra{n} \right),
\end{equation}
where $\ket{n}$ are the energy eigenstates.  For instance, if the initial state of the
system is an arbitrary ensemble of energy eigenstates
\begin{equation} \label{eq:rho0}
  \op{\rho}_0=\sum_{n=1}^N w_n \ket{n}\bra{n},
\end{equation}
where $w_n$ is the initial population of state $\ket{n}$ satisfying $0\le w_n\le 1$ 
and $\sum_{n=1}^N w_n=1$, then we have
\begin{eqnarray}
 \op{U}_2\op{\rho}_0\op{U}_2^\dagger 
 &=& \op{U}_1 \sum_{n=1}^N \ket{n} e^{i\theta_n} w_n e^{-i\theta_n} \bra{n} 
     \op{U}_1^\dagger\nonumber\\
 &=& \op{U}_1\sum_{n=1}^N w_n\ket{n}\bra{n}\op{U}_1^\dagger 
      =\op{U}_1\op{\rho}_0\op{U}_1^\dagger, 
\end{eqnarray}
i.e., the phase factors $e^{i\theta_n}$ cancel out.  Thus, if the initial state of the
system is an ensemble of energy eigenstates, which of course includes trivial ensembles
such as pure energy eigenstates, then we only need to find a Lie group decomposition of
the target operator $\op{U}$ modulo phase factors, i.e., it suffices to find matrices 
$\op{V}_k$ such that
\begin{equation}
  \op{U} \left(\sum_{n=1}^N e^{i\theta_n} \ket{n}\bra{n} \right)
         = \op{V}_K \op{V}_{K-1} \cdots \op{V}_1.
\end{equation}
Note that decomposition modulo phase factors, when sufficient, is more efficient since
it requires in general up to $2(N-1)$ fewer steps than the general decomposition
algorithm.  See appendix \ref{appendix:Udecomp} for details.

\section{Amplitude and pulse length}
\label{sec:amp}

Comparing equations (\ref{eq:Vk1}) and (\ref{eq:Vk2}) shows that the constants $C_k$
in the decomposition (\ref{eq:Udecomp}) determine the pulse area of the $k$th pulse:
\begin{equation} \label{eq:pulsearea}
   \int_{t_{k-1}}^{t_k} \!\!\!\!\! 2 A_k(\tau)\,d\tau 
   = 2\int_{t_{k-1}}^{t_k} \!\!\!\!\! A_k(\tau)\,d\tau 
   \equiv\frac{2 C_k}{d_{\sigma(k)}}.
\end{equation}
There is considerable flexibility as regards the shape of the control pulses that are 
used.  In this paper we consider two simple types of control functions:  square-wave
pulses with finite rise and decay times and Gaussian wavepackets.  The former type of
control pulses is convenient since it is easy to realize them the laboratory.  In the
optical regime, for instance, a combination of CW lasers and Pockel cells can be used
to achieve such pulse shapes.  Gaussian control pulses can be derived from pulse laser
systems.  In addition, the latter type of control pulses has the distinct advantage of
minimal frequency dispersion, which is generally desirable for frequency-selective
control.

\subsection{Square-wave pulses}
\label{subsec:SWP}

If we apply a field of constant amplitude $2A_k$ for a fixed period of time $\Delta t_k
=t_k-t_{k-1}$ then $A_k$ is determined by the pulse area constraint (\ref{eq:pulsearea}) 
\begin{equation} \label{eq:Ak1}
  A_k \equiv \frac{C_k}{d_{\sigma(k)} \Delta t_k}.
\end{equation}
Thus, $A_k$ can be adjusted by changing the pulse length $\Delta t_k$, which allows us 
to account for laboratory constraints on the strengths of the control fields and limit 
undesirable off-resonant effects by ensuring that 
\begin{equation}  
   A_k \ll \min_n |\mu_{n-1}-\mu_n|/d_{\sigma(k)}.
\end{equation}

\begin{figure}
\epsfysize 2.5 in \epsfbox{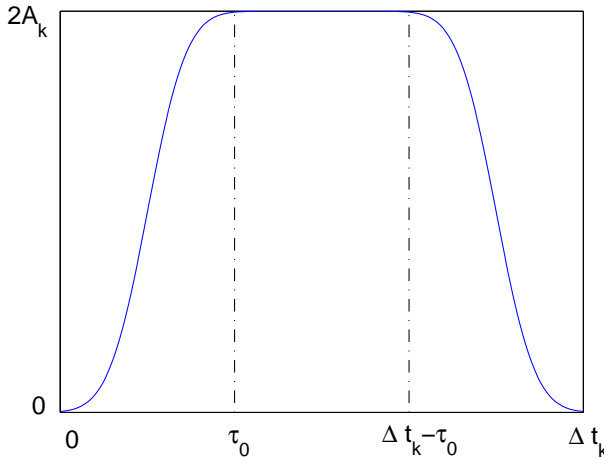}
\caption{Square wave pulse of total pulse length $\Delta t_k$ with rise and decay time 
$\tau_0$ and peak amplitude $2A_k$}\label{Fig:pulse1}
\end{figure}

Ideal square-wave pulses are rather simple and convenient pulse shapes.  However, in 
practice it is impossible to reproduce square-wave pulses exactly.  Rather, each pulse
will have a certain rise and decay time $\tau_0$, which leads to pulse envelopes of the
form depicted in Fig.~\ref{Fig:pulse1}.  Mathematically, such pulses can be modeled by
choosing the envelope of the pulse of the form
\begin{eqnarray}
  2 A_k(t) 
  &=& A_k\left\{1+\mbox{erf}\left[4(t-\tau_0/2)/\tau_0\right]\right\}\nonumber\\
  & &+A_k\left\{1+\mbox{erf}\left[4(t-\Delta t+\tau_0/2)/\tau_0\right]\right\} 
\end{eqnarray}
where $\mbox{erf}(x)$ is the error function
\begin{equation}
   \mbox{erf}(x) = \frac{2}{\sqrt{\pi}} \int_0^x \!\!\! e^{-t^2} \, dt.
\end{equation}
Although this envelope function may appear quite complicated, it turns out that its
pulse area is rather simple to compute.  Observing that the area
\[
  \Pi_1 \equiv A_k \int_0^{\tau_0/2} 1 + 
     \mbox{erf}\left[\frac{4}{\tau_0}\left(t-\frac{\tau_0}{2}\right)\right]\, dt 
\]
is equal to the area
\begin{eqnarray*}
  \Pi_2 &\equiv& A_k \int_{\tau_0/2}^{\tau_0} 2-\left\{ 1 + \mbox{erf}\left[\frac{4}{\tau_0}
                \left( t-\frac{\tau_0}{2}\right)\right] \right\}\, dt \\
        &=& A_k \int_{\tau_0/2}^{\tau_0} 1 - \mbox{erf}\left[\frac{4}{\tau_0} 
                \left(t-\frac{\tau_0}{2} \right) \right] \, dt \\
        &=& A_k \int_{\tau_0/2}^{\tau_0} 1 + \mbox{erf}\left[\frac{4}{\tau_0}
                \left(\frac{\tau_0}{2}-t\right) \right] \, dt \\
        &=&-A_k \int_{\tau_0/2}^{0} 1 + \mbox{erf}\left\{\frac{4}{\tau_0} 
                \left[\frac{\tau_0}{2}-(\tau_0-t)\right]\right\} \, dt \\
        &=& A_k \int_0^{\tau_0/2} 1 + \mbox{erf}\left[\frac{4}{\tau_0} 
                \left(t-\frac{\tau_0}{2}\right)\right] \, dt 
\end{eqnarray*}
since $-\mbox{erf}(x)=\mbox{erf}(-x)$, we note that the total pulse area of a modified
square wave pulse of peak amplitude $2 A_k$ and total pulse length $\Delta t_k\ge 2\tau_0$
with rise and decay time $\tau_0$ is simply equal to the area of a rectangle of width 
$\Delta t_k-\tau_0$ and height $2 A_k$. Thus, the pulse area constraint (\ref{eq:pulsearea})
leads to $2 A_k(\Delta t_k -\tau_0) \equiv 2C_k/d_{\sigma(k)}$ or equivalently
\begin{equation}\label{eq:Ak2}
   A_k \equiv \frac{C_k}{d_{\sigma(k)} (\Delta t_k -\tau_0)}.
\end{equation}
Although this formula is very similar to (\ref{eq:Ak1}), note the importance of including
the rise and decay time of the pulse.  Neglecting $\tau_0$ amounts to overestimating the 
pulse area, which will result in poor control.  

Unless restrictions on the pulse strength dictate otherwise, the pulse lengths $\Delta t_k$
can be chosen to be the same.  For instance, if the target time for achieving the control 
objective is $T$ and the Lie group decomposition shows that $K$ pulses are required to 
achieve the control objective, then we would usually set
\begin{equation}
  \Delta t_k = \frac{T}{K}, \quad 1\le k\le K.
\end{equation}
However, recall that it is important to assure that equation (\ref{eq:detuning1}) is
satisfied, i.e., that $2C_k/(\Delta t_k-\tau_0)$ is much smaller than the detuning of the
pulse frequency from off-resonant transitions.  For instance, for a Morse oscillator with
anharmonicity $\alpha$, the detuning of each pulse from off-resonant transitions is at 
least $2\alpha$ and thus $\Delta t_k-\tau_0$ should be much larger than $C_k/\alpha$ in
this case.

\subsection{Gaussian wavepackets}
\label{subsec:GWP}

If we wish to use Gaussian wavepackets instead of square waves then we choose the
envelope of the $k$th control pulse to be of the form
\begin{equation}
  2A_k(t) = 2A_k e^{-q_k^2 (t-\Delta t_k/2 - t_{k-1})^2},
\end{equation}
which corresponds to a Gaussian wavepacket centered at $t_k^*=t_{k-1}+\Delta t_k/2$ 
of peak amplitude $2A_k$.  The peak amplitude $2A_k$ is determined by the pulse area 
constraint (\ref{eq:pulsearea}).  Concretely, we have
\begin{equation}
 A_k \equiv \frac{q_k C_k}{d_{\sigma(k)} \sqrt{\pi}}
\end{equation}
provided that the $k$th time interval $\Delta t_k$ is large enough to justify the
assumption
\[
  \int_{t_{k-1}}^{t_k} \!\!\! e^{-q_k^2 (t-\Delta t_k/2 - t_{k-1})^2}\, dt
 \approx \int_{-\infty}^{+\infty} \!\!\! e^{-q^2 \tau^2} \, d\tau 
  = \frac{\sqrt{\pi}}{q_k}.  
\]
Noting that
\[
 \int_{-\Delta t_k/2}^{\Delta t_k/2} \!\!\! e^{-q_k^2 t^2}\, dt
 = \frac{\sqrt{\pi}}{q_k} \mbox{erf}(q_k \Delta t_k/2)
\]
and $\mbox{erf}(2)=0.995322$ we see, for instance, that choosing 
\begin{equation}
  q_k = \frac{4}{\Delta t_k}
\end{equation}
guarantees that more than 99\% of the pulse area of the $k$th pulse is contained in the
control interval $[t_{k-1},t_k]$.  Choosing $q_k=6/\Delta t_k$ would ensure that more than
$99.99$\% of the pulse area of the $k$th pulse is contained in the $k$th control interval.
Again, unless restrictions on the pulse strength dictate otherwise, we set
\begin{equation}
  \Delta t_k = \frac{T}{K}, \quad 1\le k\le K,
\end{equation}
where $T$ is the target time and $K$ is the number of pulses required.  Note that the peak
amplitude $2A_k$ of the $k$th field is proportional to $q_k$, which is in turn inversely 
proportional to $\Delta t_k$.  Hence, we can again limit the peak amplitudes and thus 
off-resonant effects by choosing the pulse lengths $\Delta t_k$ sufficiently large.  
\begin{figure}
\epsfysize 2.5 in \epsfbox{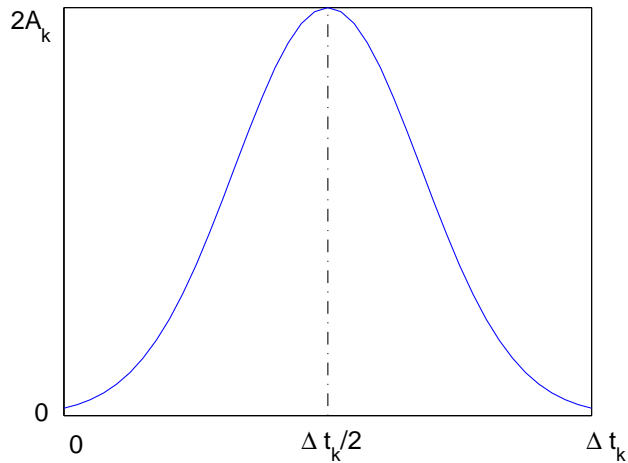}
\caption{Gaussian wavepacket with $q_k=\frac{4}{\Delta t_k}$ with peak amplitude $2A_k$}
\label{Fig:pulse2}
\end{figure}
In the following sections we apply these results to various control problems.  

\section{Population transfer $\ket{1}\rightarrow\ket{N}$ for a $N$-level system}
\label{sec:poptransfer}

Let us first consider a $N$-level system initially in the ground state $\ket{1}$ and
apply the decomposition technique to the simple problem of transferring the population
of state $\ket{1}$ to the highest excited state $\ket{N}$ by applying a sequence of 
monochromatic control pulses, each resonant with one of the transitions frequencies 
$\mu_m$, which can be regarded as the population inversion route to short-wavelength 
lasers.  It can easily be verified that any evolution operator $\op{U}$ of the form
\begin{equation} \label{eq:U1}
  \op{U} = \left( \begin{array}{c|c}
            \vec{0}       & \; A_{N-1} \\\hline
            e^{i\theta}\; & \; \vec{0}
           \end{array} \right),
\end{equation}
where $A_{N-1}$ is an arbitrary unitary $(N-1)\times (N-1)$ matrix, $e^{i\theta}$ is 
an arbitrary phase factor and $\vec{0}$ is a $N-1$ vector whose elements are $0$, 
achieves the control objective since
\[
  \left(\begin{array}{c|c}
            \vec{0}       & \; A_{N-1} \\\hline
            e^{i\theta}\; & \; \vec{0}
          \end{array}\right) 
   \left( \begin{array}{c} 1       \\ \vec{0}       \end{array} \right)
 = \left( \begin{array}{c} \vec{0} \\ e^{i\theta_N} \end{array} \right)
\]
and thus the population of state $\ket{N}$ is equal to $\sqrt{e^{-i\theta_N} 
e^{i\theta_N}}=1$ after application of $\op{U}$.  Next we observe that setting 
\begin{equation}\label{eq:Udecomp1}
   \op{U}  = \op{U}_0(T)\op{U}_I, \quad
   \op{U}_I=\op{V}_{N-1} V_{N-2} \cdots V_1,
\end{equation}
where the factors are
\begin{eqnarray} \label{eq:Vm}
 \op{V}_m &=& \exp\left[ \frac{\pi}{2}
              \left(\sin\phi_m\op{x}_m -\cos\phi_m \op{y}_m\right)\right] \\
          &=& -i(e^{i\phi_m} \op{e}_{m,m+1}+e^{-i\phi_m}\op{e}_{m+1,m})
               + \sum_{n\neq m,\atop n\neq m+1} \op{e}_{n,n} \nonumber
\end{eqnarray}
for $1\le m\le N-1$, always leads to a $\op{U}$ of the form (\ref{eq:U1}), independent 
of the initial pulse phases $\phi_m$.  This factorization corresponds to a sequence of 
$N-1$ control pulses, where the $m$th pulse is resonant with the transition $\ket{m}
\rightarrow\ket{m+1}$ and has pulse area $\pi/d_m$, which transfers the population step
by step to the target level
\[ 
   1 \rightarrow 2 \rightarrow 3 \rightarrow \cdots \rightarrow N.
\]
Thus, the control sequence derived using the Lie group decomposition technique agrees 
in this case with the obvious choice of the control pulses.  

\begin{figure*}
\epsfbox{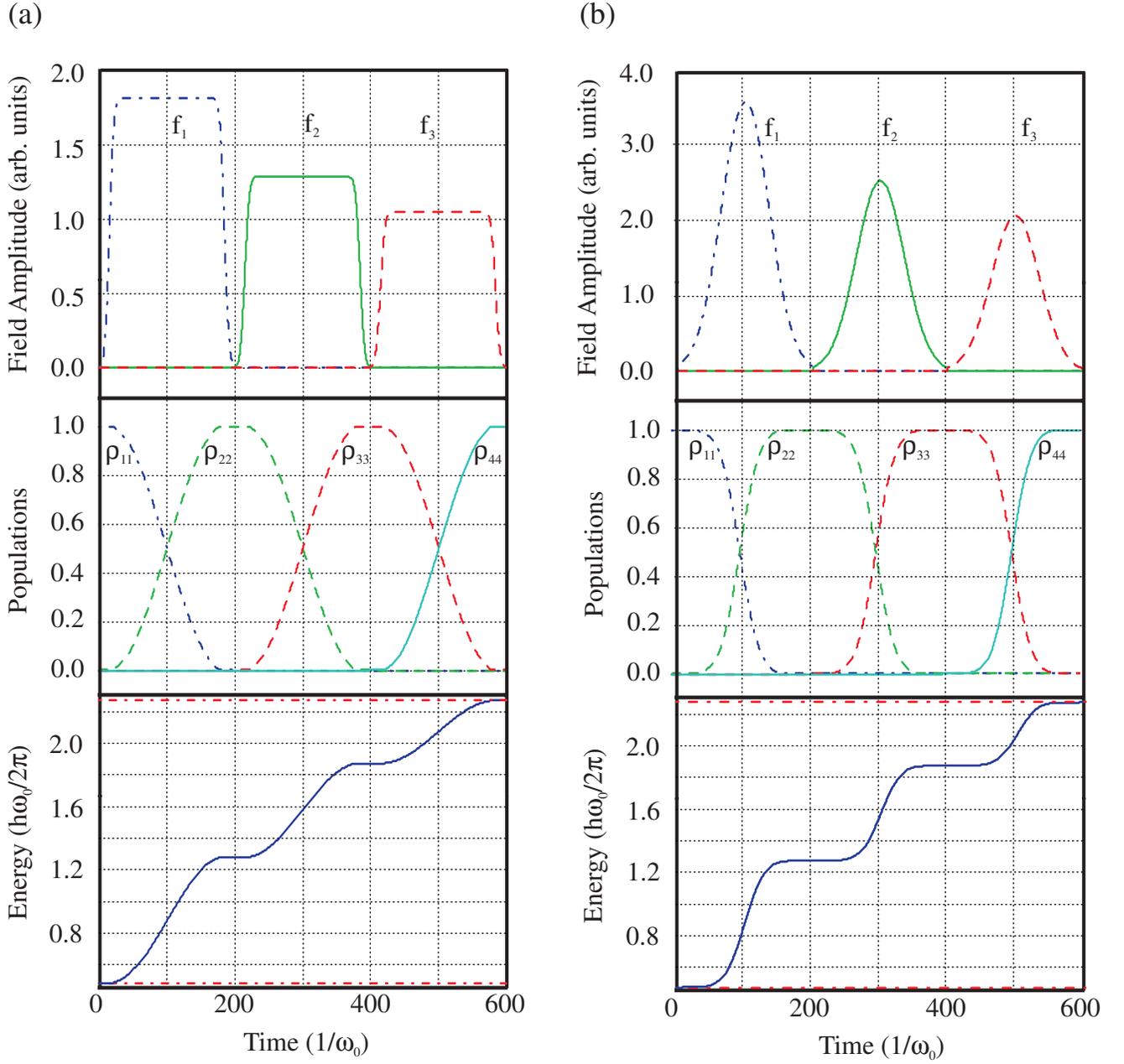}
\caption{Population transfer from the ground state $\ket{1}$ to the excited state $\ket{4}$
for the four-level Morse oscillator described above using three square-wave control pulses
with rise and decay time $\tau_0=30$ time units (a) and Gaussian pulses with shape factor 
$q=4/100$ (b).  The $m$th pulse is resonant with the transition $\ket{m}\rightarrow\ket{m+1}$
and has pulse area $\pi/d_m$ for $m=1,2,3$.}
\label{Fig:PopTransfer} 
\end{figure*}

Figure \ref{Fig:PopTransfer} shows the results of control computations for the four-level
Morse oscillator described above using square-wave and Gaussian control pulses, respectively.
The top graph in each figure shows the envelopes of the control pulses in units of $10^{-2}
\hbar\omega_0/p_{12}$, where $\omega_0$ is the oscillator frequency and $p_{12}$ is the 
dipole moment of the $1\rightarrow 2$ transition.  The labels $f_m$ for $m=1,2,3$ in the 
field plot indicate that the corresponding pulse is resonant with the transition $\ket{m}
\rightarrow\ket{m+1}$. 
The middle graphs shows the evolution of the energy-level populations resulting from the 
application of the control fields, and the bottom graphs the energy of the system as a 
function of time, where the upper and lower horizontal lines in the bottom graph indicate
the kinematical upper and lower bounds for the energy.  Observe that for both choices of 
the control pulses the final population of target level four is one, i.e., complete 
population transfer is achieved, while the populations of the intermediate levels increase
and decay intermittently.  

Also note that the energy of the system increases monotonically from its kinematical 
minimum value at $t=0$ to its maximum value at the final time in both cases.  The 
gradient of approach, however, is more uniform for square-wave pulses, while Gaussian
pulses tend to result in short, steep increases with long intermittent plateau regions.
This is an advantage of using square-wave pulses, which could be significant especially
for realistic systems where spontaneous emission is a problem.  Gaussian wavepackets on
the other hand, have the distinct advantage of minimal frequency dispersion and should
thus be less likely to induce unwanted off-resonant effects.

As regards off-resonant effects in general, note that $C_k=\pi/2$ for all pulses.  Thus,
for our Morse oscillator with $\alpha=0.1$ and square-wave pulses with rise and decay 
time $\tau_0=30$ time units, we have $C_k/\alpha=5\pi$, which is much smaller than our
choice of $\Delta t_k-\tau_0=170$ time units.  More precisely, the peak amplitude of field
$f_m$, which is resonant with the transition $\ket{m}\rightarrow\ket{m+1}$, is 
\[
  2 A_m= \frac{\pi}{(200-30) \sqrt{m}} \; \frac{\hbar\omega_0}{p_{12}}.
\] 
Recalling that the dipole moment is $d_m=\sqrt{m}$ $p_{12}$, the ``Rabi-frequency'' of 
the pulse is thus $2 A_m d_m/\hbar=\pi/170$ $\omega_0$, which is less than $1/10$ of 0.2
$\omega_0$, the detuning of the pulse frequency from closest off-resonant transitions.
Note that for fixed pulse lengths, the peak amplitudes for Gaussian control pulses are
necessarily larger than for square-wave pulses, in our case by about a factor of 1.5.
However, it seems reasonable to assume that Gaussian control pulses that have the same
pulse area and pulse length as square-wave pulses should be less likely to cause strong
off-resonant effects than square-wave pulses due to smaller frequency dispersion.

\section{Inversion of ensemble populations for a mixed-state system}
\label{sec:inversion}

The results of the previous section are encouraging in that they agree with intuitive
control schemes.  The power of the decomposition technique, however, lies in its ability to
predict control schemes for problems that have no such obvious solutions.  One such case is
a generalization the control problem discussed in the previous section, where we attempt to
achieve complete inversion of the ensemble populations for a $N$-level system whose initial
state $\op{\rho}_0$ is an arbitrary ensemble of energy eigenstates (\ref{eq:rho0}).  This 
control operation can be regarded as a form of generalized NOT gate for mixed-state $N$-level
systems, the ensemble NOT gate, which should not be confused with the U-NOT gate described 
by Buzek \emph{et.al.} in \cite{JMO47p211}.  

The desired evolution operator to achieve a complete reversal of the ensemble populations
for the system is
\begin{equation} \label{eq:U2}
  \op{U}=\left(\begin{array}{cccccc}
            0 & 0 & \cdots & 0 & e^{i\theta_1} \\
            0 & 0 & \cdots & e^{i\theta_2} & 0 \\
            \vdots& \vdots & & \vdots & \vdots \\
            0 & e^{i\theta_{N-1}} & \cdots & 0 & 0\\
            e^{i\theta_N} & 0     & \cdots & 0 & 0
          \end{array}\right),
\end{equation}
where the $e^{i\theta_n}$ are arbitrary phase factors.  Assuming as above that each 
transition between adjacent energy levels can be individually addressed by selecting 
the frequency of the control pulse and possibly by using other selection rules, the 
generators of the dynamical Lie algebra are again of the form (\ref{eq:Vk1}) and a 
possible Lie group decomposition in terms of these generators is given by
\begin{equation} \label{eq:Udecomp2}
  \op{U} = \op{U}_0(T) \prod_{\ell=N-1}^{1} \left[\prod_{m=1}^\ell \op{V}_m \right],
\end{equation}
where the factors $\op{V}_m$ are as defined in (\ref{eq:Vm}).  This decomposition 
corresponds to a sequence of $K=N(N-1)/2$ control pulses
\[
  f_1, f_2, \ldots, f_{N-1}, 
  f_1, f_2, \ldots, f_{N-2}, 
 \ldots,
  f_1, f_2,
  f_1,
\]
where the $k$th pulse is resonant with the transition $\ket{\sigma(k)}\rightarrow
\ket{\sigma(k)+1}$ and has pulse area $\pi/d_{\sigma(k)}$.  This decomposition is optimal
in the sense that a complete inversion of the ensemble populations cannot be achieved with
$K'<K$ control pulses if the initial populations $w_n$ satisfy $w_n\neq w_m$ for $n\neq m$.

\begin{figure*}
\epsfbox{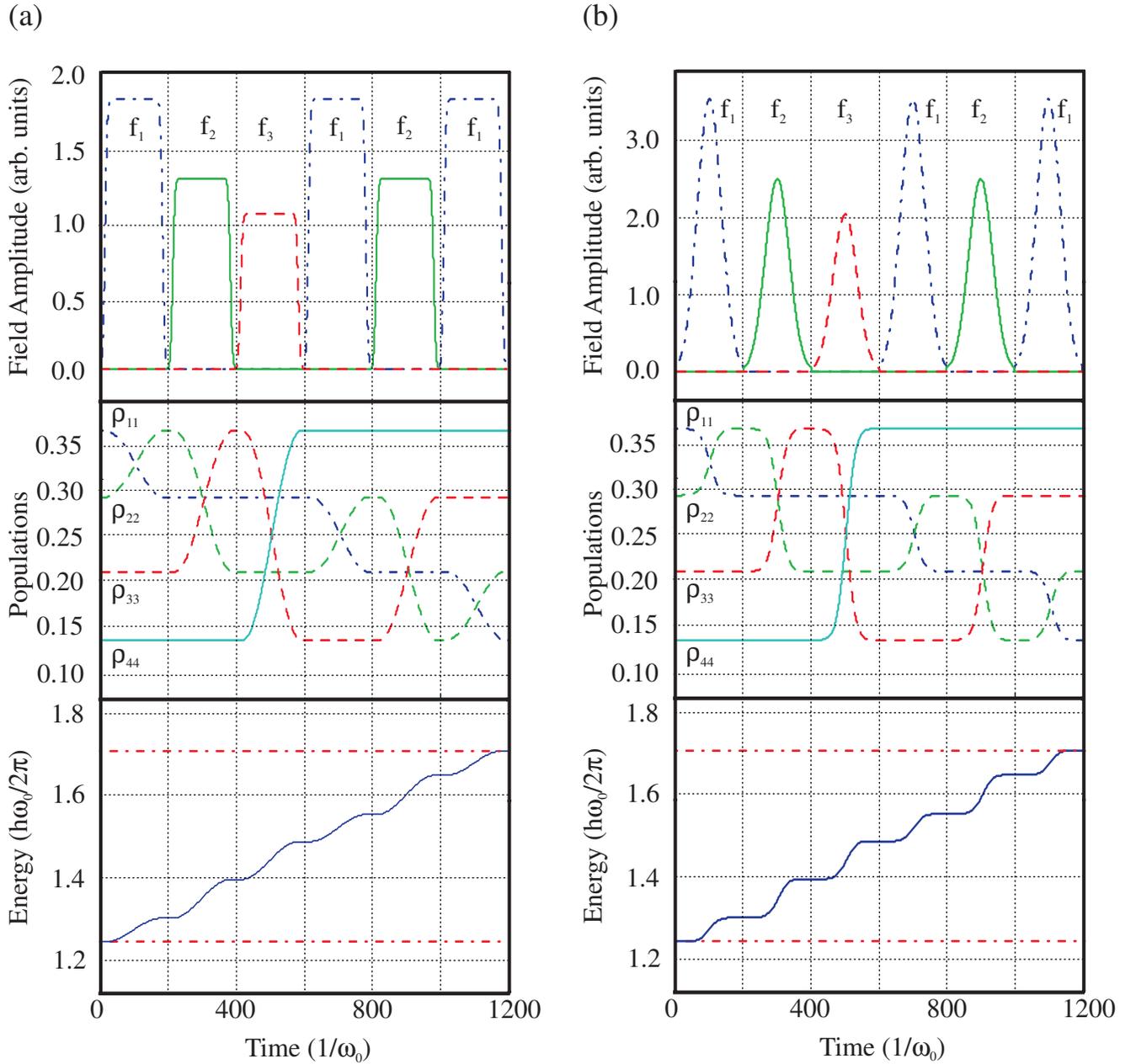}
\caption{Population inversion for a four-level Morse oscillator initially in thermal 
equilibrium using six square-wave control pulses with rise and decay time $\tau_0=30$
time units (a) and Gaussian pulses with shape factor $q=4/100$ (b).  Each pulse labeled
$f_m$ is resonant with the transition $\ket{m}\rightarrow\ket{m+1}$ and has pulse area 
$\pi/d_m$.} \label{Fig:PopInversion} 
\end{figure*}

As an example, we again apply the scheme to the $N$-level Morse oscillator with energy 
levels (\ref{eq:EnMorse}) and transition dipole moments (\ref{eq:dn}) described above.  To
be specific, we assume that the system is initially in thermal equilibrium at a temperature 
$(E_4-E_1)/k$, i.e., the initial populations of the energy eigenstates are given by the 
Boltzmann distribution 
\begin{equation}
   w_n = \frac{\exp\left[ (E_n-E_1)/(E_N-E_1) \right]}
              {\sum_{k=1}^N \exp\left[ (E_k-E_1)/(E_N-E_1) \right]}
\end{equation}
for $n=1,2,3,4$.  Note that the initial populations satisfy $w_1<w_2<w_3<w_4$.  Our goal is
to create an anti-thermal ensemble, i.e., an ensemble where the populations of the energy
eigenstates are reversed so that the ground state $\ket{1}$ has the lowest population $w_4$,
and the highest excited state has the highest population $w_1$, etc.  
Figure \ref{Fig:PopInversion} shows the result of control simulations using square-wave and 
Gaussian controls, respectively.  Note that each pulse in the control sequence interchanges
the populations of two adjacent energy levels until a complete reversal of the populations
is achieved;  for our four-level system with initial populations $w_n$ the effect of the 
controls on the populations can be summarized as follows
\[\begin{array}{l*{13}{c}}
          &     & f_1 &     & f_2 &     & f_3 &     & f_1 &     & f_2 &     & f_1 &    \\
 \ket{1}: & w_1 & \ra & w_2 & \ra & w_2 & \ra & w_2 & \ra & w_3 & \ra & w_3 & \ra & w_4\\
 \ket{2}: & w_2 & \ra & w_1 & \ra & w_3 & \ra & w_3 & \ra & w_2 & \ra & w_4 & \ra & w_3\\
 \ket{3}: & w_3 & \ra & w_3 & \ra & w_1 & \ra & w_4 & \ra & w_4 & \ra & w_2 & \ra & w_2\\
 \ket{4}: & w_4 & \ra & w_4 & \ra & w_4 & \ra & w_1 & \ra & w_1 & \ra & w_1 & \ra & w_1 
\end{array}\]
where $f_m$ refers to a control pulse of frequency $\mu_m$ and pulse area $\pi/d_m$ for 
$m=1,2,3$.  The first pulse interchanges the populations of levels $\ket{1}$ and $\ket{2}$,
the second pulse flips the populations of levels $\ket{2}$ and $\ket{3}$, the third pulse
switches the populations of levels $\ket{3}$ and $\ket{4}$, etc.

Also note that the energy of the system increases monotonically from its minimum value 
in thermal equilibrium to its kinematical maximum value at the final time, as expected,
where the gradient of approach is more uniform for square-wave pulses than for Gaussian
wavepackets.  As regards off-resonant effects, the same considerations as in the previous
section apply, since $C_k=\pi/2$ for all pulses, as in the previous example.

\section{Creation of arbitrary superposition states}
\label{sec:superposition}

In this section we consider the problem of creating arbitrary superposition states from
an initial state.  Control schemes that create such superposition states may prove very
useful in controlling quantum interference in multi-state systems and can be considered
a generalization of $\pi/2$ pulses used routinely in free induction-decay experiments. 
To be specific, we assume that the system is initially in the ground state $\ket{1}$ and
that the goal is to achieve a superposition state $\ket{\Psi(t)}$.  Observe that any 
(normalized) superposition state can be conveniently written as  
\begin{equation}\label{eq:superpos}
  \ket{\Psi(t)} = \sum_{n=1}^N r_n e^{i\theta_n} \ket{\tilde{n}(t)},
\end{equation}
where $\ket{\tilde{n}(t)}=e^{-iE_nt}\ket{n}$ is a rotating frame and the coefficients 
$r_n$ satisfy the normalization condition $\sum_{n=1}^N r_n^2=1$.  Thus, in order to 
reach the target state $\ket{\Psi(t)}$ at time $T$ we need to find a unitary operator 
$\op{U}(T)$ such that
\begin{equation} \label{eq:U-1}
    \op{U}(T) 
   \left(\begin{array}{c} 1 \\ 
                          0 \\ 
                          \vdots \\ 
                          0 
   \end{array} \right)
 = \left(\begin{array}{c} r_1 e^{i(\theta_1-E_1T)} \\ 
                          r_2 e^{i(\theta_2-E_2T)} \\ 
                          \vdots \\ 
                          r_N e^{i(\theta_N-E_NT)}
          \end{array} \right).
\end{equation}
However, note that it actually suffices to find a unitary operator $\op{U}_1$ such that
\begin{equation} \label{eq:U-2}
    \op{U}_1 \left(\begin{array}{c} 1 \\ 0 \\ \vdots \\ 0 \end{array} \right)
  = \left(\begin{array}{c} r_1 \\ r_2 \\ \vdots \\ r_N \end{array} \right)
\end{equation}
since if $\op{U}_1$ satisfies (\ref{eq:U-2}) then $\op{U}(T)=\op{\Theta}(T)\op{U}_1$ where 
\begin{equation}
  \op{\Theta}(T) \equiv \mbox{diag}(e^{i(\theta_1-E_1T)},\ldots,e^{i(\theta_N-E_NT)})
\end{equation}
automatically satisfies (\ref{eq:U-1}).  In order to find a unitary operator $\op{U}_1$
that satisfies (\ref{eq:U-2}) we set
\begin{equation}
  \hat{U}^{(0)}=\left( \begin{array}{c|c} 
                r_1 & \vec{0} \\\hline
                r_2 &  \\
             \vdots &  \op{I}_{N-1}\\
                r_N & \\ 
               \end{array} \right),
\end{equation}
where $\op{I}_{N-1}$ is the identity matrix in dimension $N-1$, and perform Gram-Schmidt
orthonormalization on the columns of $\op{U}^{(0)}$.  The resulting matrix $\op{U}_1$ 
will be unitary and satisfy (\ref{eq:U-2}).  To determine the control pulse sequence, we
then apply the decomposition algorithm presented in appendix \ref{appendix:Udecomp} to 
factor the target operator
\begin{equation}
   \op{U} = \op{U}_0(T)^\dagger\op{\Theta}(T) \op{U}_1 = \Theta(0) \op{U}_1.
\end{equation}

\begin{figure*}
\epsfbox{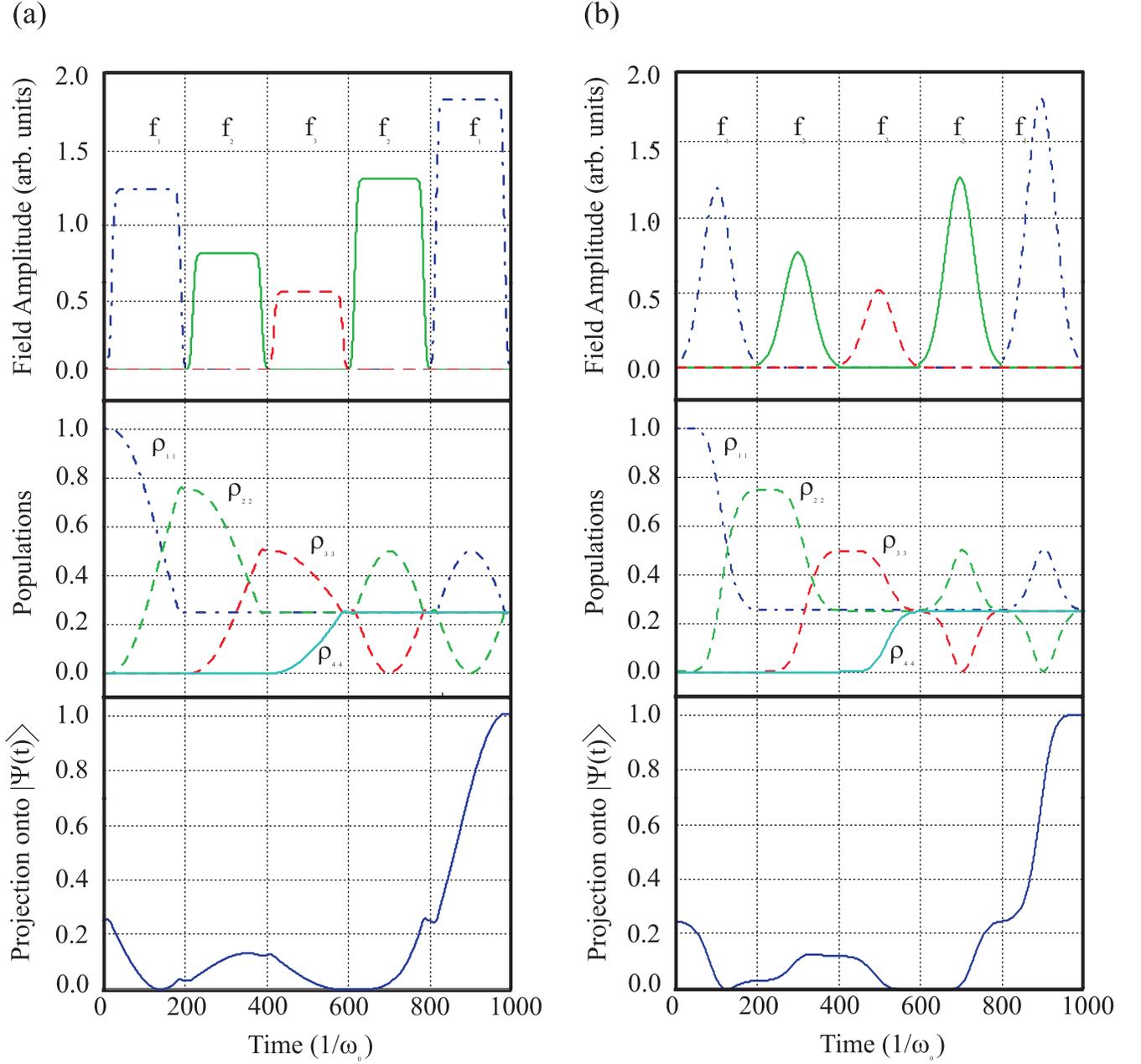}
\caption{Creation of the superposition state $\ket{\Psi(t)}=\frac{1}{2}\sum_{n=1}^4
\ket{\tilde{n}(t)}$ for a four-level Morse oscillator initially in the ground state $\ket{1}$
using square-wave control pulses with rise and decay time $\tau_0=30$ time units (a) and 
Gaussian control pulses with shape factor $q=4/100$ (b).  Each pulse labeled $f_m$ is 
resonant with the transition $\ket{m}\rightarrow\ket{m+1}$.  The pulse areas are $2\pi/3d_1$,
$2\arctan(\sqrt{2})/d_2$, $\pi/2d_3$, $\pi/d_2$ and $\pi/d_1$, respectively.}
\label{Fig:Superpos} 
\end{figure*}

As an example, let us consider the problem of creating the superposition state
\begin{equation} \label{eq:superpos1}
  \ket{\Psi(t)} = \frac{1}{2} \sum_{n=1}^4 \ket{\tilde{n}(t)}
                = \frac{1}{2} \sum_{n=1}^4 e^{-iE_n t} \ket{n}
\end{equation}
for a four-level system initially in state $\ket{1}$.  In order to find a unitary 
operator $\op{U}$ such that
\begin{equation}
   \op{U}\left(\begin{array}{c} 1 \\ 0 \\ 0 \\ 0 \end{array} \right)
 = \left(\begin{array}{c} 1/2\\ 1/2\\ 1/2\\ 1/2  \end{array} \right)
\end{equation}
we set 
\begin{equation}
  \hat{U}^{(0)}=\left(\begin{array}{cccc} 
                1/2 & 0 & 0 & 0 \\
                1/2 & 1 & 0 & 0 \\
                1/2 & 0 & 1 & 0 \\
                1/2 & 0 & 0 & 1
                \end{array} \right).
\end{equation}
and perform Gram-Schmidt orthonormalization on the columns of $\op{U}^{(0)}$, which gives
\begin{equation} \label{eq:targetU1}
  \op{U}_1=\left(\begin{array}{cccc}
   1/2 & -\sqrt{3}/6 & -\sqrt{6}/6 & -\sqrt{2}/2 \\
   1/2 & +\sqrt{3}/2 & 0           & 0 \\
   1/2 & -\sqrt{3}/6 & +\sqrt{6}/3 & 0 \\
   1/2 & -\sqrt{3}/6 & -\sqrt{6}/6 & +\sqrt{2}/2
  \end{array}\right).
\end{equation}
Since $\op{\Theta}(T)=\op{U}_0(T)$, the target operator is simply $\op{U}_0(T)^\dagger
\op{\Theta}(T)\tilde{U}=\op{U}_1$.  Applying the decomposition algorithm described in
appendix \ref{appendix:Udecomp} leads to the following factorization of $\op{U}_1$:
\begin{equation} \label{eq:Udecomp3}
  \op{U}_1 = \op{V}_5 \op{V}_4 \op{V}_3 \op{V}_2 \op{V}_1,
\end{equation}
where the factors are
\begin{equation} \label{eq:Udecomp3b}
\begin{array}{rcl}
  \op{V}_1 &=& \exp\left(+C_1\op{x}_1 \right), \quad C_1 = \frac{\pi}{3},\\
  \op{V}_2 &=& \exp\left(-C_2\op{x}_2 \right), \quad C_2 = \arctan\left(\sqrt{2}\right) \\
  \op{V}_3 &=& \exp\left(+C_3\op{x}_3 \right), \quad C_3 = \frac{\pi}{4},\\
  \op{V}_4 &=& \exp\left(+C_4\op{x}_2 \right), \quad C_4 = \frac{\pi}{2},\\
  \op{V}_5 &=& \exp\left(-C_5\op{x}_1 \right), \quad C_5 = \frac{\pi}{2}.
\end{array}
\end{equation}
This decomposition corresponds to the following sequence of five control pulses
\[\begin{array}{rll}
  f_1(t) &= A_1(t) e^{i(\mu_1 t +\pi/2)} + \mbox{c.c.} &= -2A_1(t) \sin(\mu_1 t) \\
  f_2(t) &= A_2(t) e^{i(\mu_2 t -\pi/2)} + \mbox{c.c.} &= +2A_2(t) \sin(\mu_2 t) \\ 
  f_3(t) &= A_3(t) e^{i(\mu_3 t +\pi/2)} + \mbox{c.c.} &= -2A_3(t) \sin(\mu_3 t) \\ 
  f_4(t) &= A_4(t) e^{i(\mu_2 t +\pi/2)} + \mbox{c.c.} &= -2A_4(t) \sin(\mu_2 t) \\ 
  f_5(t) &= A_5(t) e^{i(\mu_1 t -\pi/2)} + \mbox{c.c.} &= +2A_5(t) \sin(\mu_1 t) \\ 
\end{array}\]
with pulse areas $2\pi/3d_1$, $2\arctan(\sqrt{2})/d_2$, $\pi/2d_3$, $\pi/d_2$ and $\pi/d_1$,
respectively.  Note that only five instead of six pulses are required in this case since the
target operator $\op{U}_1$ has two consecutive zeros in the last column, which implies that
one of the six control pulses has zero amplitude and can thus be omitted.  The results of 
two control simulations based on this decomposition of $\op{U}_1$ using square-wave and
Gaussian control pulses, respectively, are presented in figure \ref{Fig:Superpos}.  Observe 
that the expectation value of the projector onto the target state $\ket{\Psi(t)}$ assumes 
its upper bound of one at the final time in both cases, which implies that the system has 
indeed reached the target state $\ket{\Psi(t)}$.

Unlike decompositions (\ref{eq:Udecomp1}) and (\ref{eq:Udecomp2}) in the previous sections,
in which the initial phases $\phi_m$ of the control pulses were arbitrary, the factorization
(\ref{eq:Udecomp3}) fixes the pulse area and frequency $\mu_m$ as well as the initial phase
$\phi_m$ of each pulse.  While the dependence of the control scheme on the pulse area and 
frequency is expected, the dependence on the initial phases of pulses may be a reason for 
concern.  Thus, the question arises how the initial phases of the control pulses affect the
outcome of the control process.  In order to answer this question, we compute the unitary 
operator 
\begin{equation}
  \op{U}_2 = \tilde{V}_5 \tilde{V}_4 \tilde{V}_3 \tilde{V}_2 \tilde{V}_1,
\end{equation}
where the factors are
\begin{equation}
\begin{array}{rcl}
  \tilde{V}_1 &=& \exp\left[C_1(\sin\phi_1\op{x}_1-\cos\phi_1\op{y}_1)\right], \\
  \tilde{V}_2 &=& \exp\left[C_2(\sin\phi_2\op{x}_2-\cos\phi_2\op{y}_2)\right], \\
  \tilde{V}_3 &=& \exp\left[C_3(\sin\phi_3\op{x}_3-\cos\phi_3\op{y}_3)\right], \\
  \tilde{V}_4 &=& \exp\left[C_4(\sin\phi_4\op{x}_2-\cos\phi_4\op{y}_2)\right], \\
  \tilde{V}_5 &=& \exp\left[C_5(\sin\phi_5\op{x}_1-\cos\phi_5\op{y}_1)\right]
\end{array}
\end{equation}
and the constants $C_k$ are as in (\ref{eq:Udecomp3b}) but the initial phases $\phi_k$ of
the control pulses are allowed to be arbitrary, and apply this operator to the initial 
state $\ket{1}$.  The resulting state 
\begin{equation}
  \op{U}_2 \left(\begin{array}{c} 1 \\ 0 \\ 0 \\ 0 \end{array} \right)
= \left(\begin{array}{l} 1/2 \; e^{i(\phi_4+\phi_5-\phi_1-\phi_2)} \\ 
                         1/2 \; e^{i(-\pi/2-\phi_5)} \\ 
                         1/2 \; e^{i(\pi-\phi_1-\phi_4)} \\ 
                         1/2 \; e^{i(\pi/2-\phi_1-\phi_2-\phi_3)} 
         \end{array} \right).
\end{equation}
differs from the desired target state only in the phase factors, i.e., the pulse phases do
not affect the relative amplitudes $r_n$ of the superposition state created.  Furthermore,
we can make the phase factors equal one by choosing $\phi_2=\pi/2-2\phi_1$, $\phi_3=\phi_1$,
$\phi_4=\pi-\phi_1$ and $\phi_5=-\pi/2$, where $\phi_1$ can be arbitrary.  Note that the 
choice $\phi_1=\pi/2$, $\phi_2=-\pi/2$, $\phi_3=\pi/2$, $\phi_4=\pi/2$ and $\phi_5=-\pi/2$
we made above satisfies these conditions.  Recalling that we expanded the state with respect
to a rotating frame $\ket{\tilde{n}(t)}=e^{-iE_nt}\ket{n}$, shows that the phases associated
with the energy eigenstates in this superposition state are oscillating rapidly compared to
the length of the control pulses.  It is therefore hardly surprising that control of relative
phases between the energy eigenstates in the superposition state requires control over the 
pulse phases.

\section{Optimization of observables}
\label{sec:optimization}

Finally, we address the problem of maximizing the ensemble average of an arbitrary 
observable for a system whose initial state is an ensemble of energy eigenstates 
(\ref{eq:rho0}).  Let us first consider the case of a time-independent observable 
$\op{A}$.  To determine the target operator required to maximize the ensemble average
$\ave{\op{A}}$ of $\op{A}$ we observe that $\ave{\op{A}}$ is bounded above by the 
kinematical upper bound \cite{PRA58p2684}
\begin{equation}
  \ave{\op{A}} \le \sum_{n=1}^N w_{\sigma(n)} \lambda_n, 
\end{equation}
where $\lambda_n$ are the eigenvalues of $\op{A}$ counted with multiplicity and ordered
in a non-increasing sequence
\begin{equation}
   \lambda_1 \ge \lambda_2 \ge \cdots \ge \lambda_N, 
\end{equation}
and $w_n$ are the populations of the energy levels $E_n$ of the initial ensemble and
$\sigma$ is a permutation of $\{1,\ldots,N\}$ such that
\begin{equation}
   w_{\sigma(1)} \ge w_{\sigma(2)} \ge \cdots \ge w_{\sigma(N)}.
\end{equation}
Observe that this universal upper bound for the ensemble average of any observable 
$\op{A}$ is dynamically attainable since the systems considered in this paper are 
completely controllable \cite{PRA63n025403,PRA63n063410}.  

Let $\ket{\Psi_n}$ for $1\le n\le N$ denote the normalized eigenstates of $\op{A}$ 
satisfying $\op{A}\ket{\Psi_n}=\lambda_n \ket{\Psi_n}$ and let $\op{U}_1$ be a unitary 
transformation such that
\begin{equation} \label{eq:U1def}
    \ket{\Psi_{\sigma(n)}}=\op{U}_1\ket{n}, \quad 1\le n \le N.
\end{equation}
Given an initial state $\op{\rho}_0$ of the form (\ref{eq:rho0}), we have
\begin{eqnarray}
  \Tr \left(\op{A}\op{U}_1\rho_0 \op{U}_1^\dagger\right) 
  &=& \Tr\left(\op{A}\sum_n w_n \op{U}_1\ket{n}\bra{n}\op{U}_1^\dagger\right) 
       \nonumber\\
  &=& \Tr\left(\sum_n w_n \op{A}\ket{\Psi_{\sigma(n)}}\bra{\Psi_{\sigma(n)}}\right)
       \nonumber\\
  &=& \Tr\left(\sum_n w_n\lambda_{\sigma(n)}\ket{\Psi_{\sigma(n)}}\bra{\Psi_{\sigma(n)}}\right)\nonumber\\
  &=& \sum_n w_n \lambda_{\sigma(n)} = \sum_n w_{\sigma(n)} \lambda_n.
\end{eqnarray}
Hence, if the system is initially in the state (\ref{eq:rho0}) then $\op{U}_1$ is a target
operator for which the observable $\op{A}$ assumes its kinematical maximum and we can use
the decomposition algorithm described in appendix \ref{appendix:Udecomp} to obtain the 
required factorization of the operator $\op{U}_I=\op{U}_0(T)^\dagger\op{U}_1$.

However, if $\op{A}$ is an observable whose eigenstates are not energy eigenstates, then
the expectation value or ensemble average of $\op{A}$ will usually oscillate rapidly as 
a result of the action of the free evolution operator $\op{U}_0(t)$.  These oscillations
are rarely significant for the application at hand and often distracting.  In such cases 
it is advantageous to define a dynamic observable
\begin{equation} 
  \tilde{A}(t) = \op{U}_0(t) \op{A} \op{U}_0(t)^\dagger
\end{equation}
and optimize its ensemble average instead. To accomplish this, note that if $\ket{\Psi_n}$
are the eigenstates of $\op{A}$ satisfying $\op{A}\ket{\Psi_n}=\lambda_n \ket{\Psi_n}$ 
then $\ket{\tilde{\Psi}_n(t)}=\op{U}(t)\ket{\Psi_n}$ are the corresponding eigenstates
of $\tilde{A}(t)$ since
\begin{eqnarray*}
   \tilde{A}(t) \ket{\tilde{\Psi}_n(t)}
 &=& \op{U}(t)\op{A} \op{U}(t)^\dagger \op{U}(t) \ket{\Psi_n}\\
 &=& \op{U}(t) \lambda_n \ket{\Psi_n} \\
 &=& \lambda_n \ket{\tilde{\Psi}_n(t)} 
\end{eqnarray*}
and if $\op{U}_1$ is a unitary operator such that equation (\ref{eq:U1def}) holds then
$\op{U}_0(t)\op{U}_1$ is a unitary operator that maps the energy eigenstates $\ket{n}$
onto the $\tilde{A}(t)$-eigenstates $\ket{\tilde{\Psi}_n(t)}$ since
\[
   \op{U}_0(t) \op{U}_1 \ket{n}
 = \op{U}_0(t) \ket{\Psi_{\sigma(n)}}
 = \ket{\tilde{\Psi}_{\sigma(n)}(t)} 
\]
for $1\le n\le N$.  Thus, the evolution operator required to maximize the ensemble 
average of $\tilde{A}(t)$ at time $T>0$ is $\op{U}_0(T) \op{U}_1$ and the target 
operator to be decomposed as described in appendix \ref{appendix:Udecomp} is 
\begin{equation}
  \op{U}=\op{U}_0(T)^\dagger\op{U}_0(T)\op{U}_1=\op{U}_1.
\end{equation}

For instance, suppose we wish to maximize the ensemble average of the transition dipole 
moment operator $\tilde{A}(t)=\op{U}_0(t)\op{A}\op{U}_0(t)^\dagger$, where
\begin{equation}\label{eq:A}
  \op{A}= \sum_{n=1}^{N-1} d_n \left(\ket{n}\bra{n+1}+\ket{n+1}\bra{n}\right),
\end{equation}
for a system initially in state (\ref{eq:rho0}) with 
\begin{equation}
  w_1 > w_2 > \cdots > w_N > 0.
\end{equation}
Then we need to find a unitary operator $\op{U}_1$ that maps the initial state $\ket{n}$ 
onto the $\op{A}$-eigenstate $\ket{\Psi_n}$ for $1\le n\le N$.  To accomplish this we let
$\op{U}_1$ be the $N\times N$ matrix whose $n$th column is the normalized $\op{A}$-eigenstate
$\ket{\Psi_n}$.  Then $\op{U}_1$ clearly satisfies $\op{U}_1\ket{n}=\ket{\Psi_n}$ and since
$\ip{\Psi_n}{\Psi_m}=\delta_{n,m}$ by hypothesis, i.e., the eigenstates are orthonormal,
$\op{U}_1$ is automatically unitary.

For $N=4$ and $d_n=\sqrt{n}$  the eigenvalues of the operator $\op{A}$ defined in equation 
(\ref{eq:A}) are (in decreasing order) 
\[
 \lambda_1=\sqrt{3+\sqrt{6}}, \;
 \lambda_2=\sqrt{3-\sqrt{6}}, \;
 \lambda_3=-\lambda_2, \;
 \lambda_4=-\lambda_1
\] 
and the corresponding eigenstates with respect to the standard basis $\ket{n}$ are the
columns of the operator 
\begin{equation}
 \op{U}_1=\left[ \begin{array}{cccc} 
 \frac{1}{2\lambda_1}&\frac{1}{2\lambda_2}&\frac{1}{2\lambda_2}&\frac{1}{2\lambda_1}\\[1.ex]
\frac{1}{2}         &\frac{1}{2}         &-\frac{1}{2}        &-\frac{1}{2} \\[1.ex]
 \frac{\sqrt{2}+\sqrt{3}}{2\lambda_1} & \frac{\sqrt{2}-\sqrt{3}}{2\lambda_2} &
 \frac{\sqrt{2}-\sqrt{3}}{2\lambda_2} & \frac{\sqrt{2}+\sqrt{3}}{2\lambda_1} \\[1.ex]
 \frac{1}{2}         &-\frac{1}{2}         &\frac{1}{2}        &-\frac{1}{2} \\
\end{array} \right].
\end{equation}

Applying the decomposition algorithm described in appendix \ref{appendix:Udecomp} yields 
the following factorization
\begin{equation} \label{eq:Udecomp4}
 \op{U}_1 \op{\Theta} = \op{V}_6 \op{V}_5 \op{V}_4 \op{V}_3 \op{V}_2 \op{V}_1,
\end{equation}
where the factors are
\begin{equation}\begin{array}{ll}
  \op{V}_1 = \exp\left(-C_1\op{x}_1 \right), & C_1 = \pi/4,\\
  \op{V}_2 = \exp\left(-C_2\op{x}_2 \right), & C_2 = \arctan\left(\sqrt{2}\right),\\
  \op{V}_3 = \exp\left(-C_3\op{x}_1 \right), & C_3 = 
                       \mbox{arccot}\left(\frac{\sqrt{6}-\sqrt{3}+3\sqrt{2}}{3}\right),\\
  \op{V}_4 = \exp\left(-C_4\op{x}_3 \right), & C_4 = \pi/3,\\
  \op{V}_5 = \exp\left(-C_5\op{x}_2 \right), & C_5 = 
                       \arctan\left(\frac{\sqrt{4+\sqrt{6}}}{\sqrt{2}+\sqrt{3}}\right),\\
  \op{V}_6 = \exp\left(-C_6\op{x}_1 \right), & C_5 = 
                        \mbox{arccot}\left(\sqrt{3+\sqrt{6}}\right)
\end{array}
\end{equation}
and $\op{\Theta}=\mbox{diag}(1,-1,1,-1)$.  Note that $\op{U}_2\equiv\op{U}_1\op{\Theta}$
is equivalent to $\op{U}_1$ since $\op{\Theta}$ commutes with $\op{\rho}_0$ as defined in
equation (\ref{eq:rho0}), i.e., $\op{\Theta}\op{\rho}_0\op{\Theta}^\dagger=\op{\rho}_0$,
and thus
\begin{eqnarray}
 \Tr\left(\op{A}\op{U}_2\op{\rho}_0\op{U}_2^\dagger\right) 
 &=&\Tr\left(\op{A}\op{U}_1\op{\Theta}\op{\rho}_0\op{\Theta}^\dagger\op{U}_1^\dagger\right)
 \nonumber\\
 &=&\Tr\left(\op{A}\op{U}_1 \op{\rho}_0\op{U}_1\right). \label{eq:Theta-equiv}
\end{eqnarray}

Decomposition (\ref{eq:Udecomp4}) corresponds to the following sequence of six control pulses
\[\begin{array}{rll}
  f_1(t) &= A_1(t) e^{i(\mu_1 t -\pi/2)} + \mbox{c.c.} &= 2A_1(t) \sin(\mu_1 t) \\
  f_2(t) &= A_2(t) e^{i(\mu_2 t -\pi/2)} + \mbox{c.c.} &= 2A_2(t) \sin(\mu_2 t) \\ 
  f_3(t) &= A_3(t) e^{i(\mu_1 t -\pi/2)} + \mbox{c.c.} &= 2A_3(t) \sin(\mu_1 t) \\ 
  f_4(t) &= A_4(t) e^{i(\mu_3 t -\pi/2)} + \mbox{c.c.} &= 2A_4(t) \sin(\mu_3 t) \\ 
  f_5(t) &= A_5(t) e^{i(\mu_2 t -\pi/2)} + \mbox{c.c.} &= 2A_5(t) \sin(\mu_2 t) \\ 
  f_6(t) &= A_6(t) e^{i(\mu_1 t -\pi/2)} + \mbox{c.c.} &= 2A_6(t) \sin(\mu_1 t)  
\end{array}\]
with pulse areas $\pi/2d_1$, $2C_2/d_2$, $2C_3/d_1$, $2\pi/3d_3$, $2C_5/d_2$, $2C_6/d_1$, 
respectively.  Again, the decomposition fixes the frequency and pulse area as well as 
the initial phase of each pulse, and the question thus arises, what role the phases play.
As we have already seen above, the target operator $\op{U}_1$ is not unique.  In fact,
equation (\ref{eq:Theta-equiv}) shows that right multiplication of $\op{U}_1$ by any 
unitary matrix that commutes with $\op{\rho}_0$ produces another unitary operator that 
is equivalent to $\op{U}_1$ in that both evolution operators lead to the same ensemble
average of the target observable.  Nevertheless, in general, the control process is 
sensitive to the phases $\phi_m$.  For instance, one can verify that changing the phase
$\phi_1$ of the first pulse from $-\pi/2$ to $\pi/2$ in the pulse sequence above leads to
the following evolution operator
\begin{equation}
 \op{U}_3 = \left[ \begin{array}{cccc} 
 \frac{1}{2\lambda_2}&\frac{1}{2\lambda_1}&\frac{1}{2\lambda_2}&-\frac{1}{2\lambda_1}\\[1ex]
 \frac{1}{2}         &\frac{1}{2}         &-\frac{1}{2}        &\frac{1}{2} \\[1ex]
 \frac{\sqrt{2}-\sqrt{3}}{2\lambda_2} & \frac{\sqrt{2}+\sqrt{3}}{2\lambda_1} & 
 \frac{\sqrt{2}-\sqrt{3}}{2\lambda_2} & \frac{\sqrt{2}+\sqrt{3}}{-2\lambda_1} \\[1ex]
 -\frac{1}{2}        &\frac{1}{2}     & \frac{1}{2}            & \frac{1}{2} 
\end{array} \right],
\end{equation}
which maps $\ket{3}$ onto $\ket{\Psi_3}$ and $\ket{4}$ onto $-\ket{\Psi_4}$ but $\ket{1}$ 
onto $\ket{\Psi_2}$ and $\ket{2}$ onto $\ket{\Psi_1}$ and thus leads to the ensemble average
\begin{equation}
 \ave{\op{A}} = w_1 \lambda_2 + w_2 \lambda_1 + w_3 \lambda_3 + w_4 \lambda_4
\end{equation}
at the final time, which is strictly less than the kinematical maximimum if $w_1>w_2$.

\begin{figure*}
\epsfbox{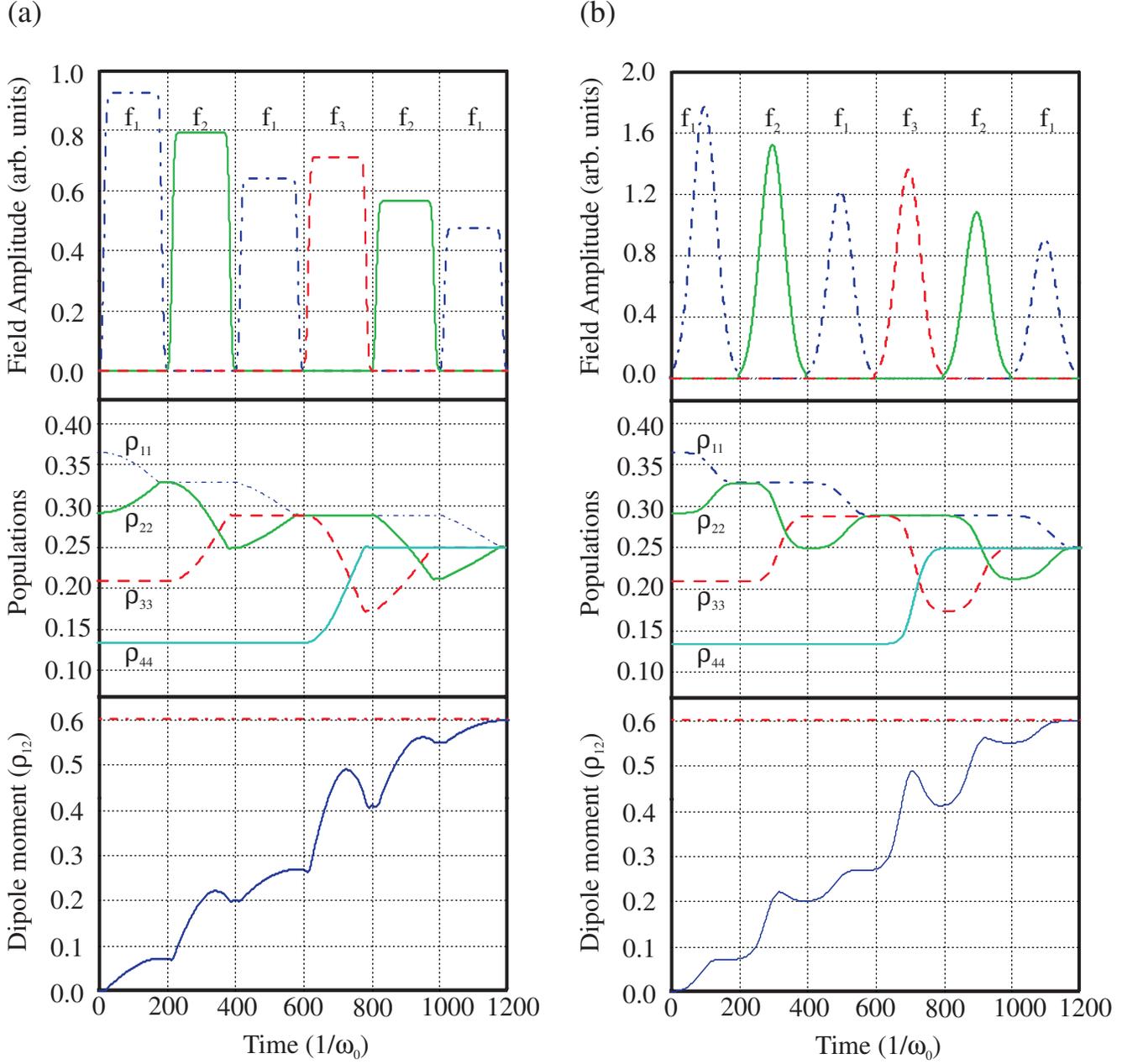}
\caption{Maximization of the (dynamic) transition dipole moment operator $\tilde{A}(t)$
for a four-level Morse oscillator initially in thermal equilibrium using square-wave 
control pulses with rise and decay time $\tau_0=30$ time units (a) and Gaussian control
pulses with shape factor $q=4/100$ (b).}\label{Fig:Dipole} 
\end{figure*}

Figure \ref{Fig:Dipole} shows the results of control simulations based on the control pulse
sequence above for our four-level Morse oscillator initially in thermal equilibrium using 
square-wave and Gaussian control pulses, respectively.  Observe that the observable indeed
attains its kinematical upper bound at the final time, as desired.

\section{Conclusion}
\label{sec:conclusion}

We have presented several control schemes designed to achieve a wide variety of control
objectives ranging from population transfers and inversions of ensemble populations to 
creation of superposition states and optimization of observables.  The main advantage of
the schemes is their simplicity: the control objective is achieved by applying a sequence
of simple control pulses such as square-wave pulses with finite rise and decay times, 
which can be created in the laboratory by modulating the amplitude of CW lasers using 
Pockel cells, or Gaussian wavepackets, which can be obtained from pulsed laser sources. 
The main shortcoming of this control approach is the need to be able to address a single
transition at a time, which limits the applicability of this technique to systems with 
sufficiently distinct transition frequencies (as regards the controlled transitions) to
permit frequency-selective control, unless other selection rules can be employed to 
address transitions individually.  Another possible complication arises from unwanted 
off-resonant effects.  However, as we have demonstrated, in general it is possible to 
limit such effects by using sufficiently weak control pulses.

\appendix 
\begin{widetext}
\section{Derivation of equation (\ref{eq:Omega})}
\label{appendix:A}

Inserting equations (\ref{eq:U0}) and (\ref{eq:Hm}) into equation (\ref{eq:SE2}) and 
recalling $\mu_m=E_{m+1}-E_m$ leads to
\begin{eqnarray*}
 i\frac{d\op{U}_I(t)}{dt} 
&=&\op{U}_0(t)^\dagger\left\{\sum_{m=1}^M\op{H}_m[f_m(t)]\right\}\op{U}_0(t)\op{U}_I(t)\\
&=&\sum_{n,m,n'} e^{i E_n t} \op{e}_{n,n} 
   \left(  A_m(t) e^{ i(\mu_m t + \phi_m)} d_m \op{e}_{m,m+1} 
           + A_m(t) e^{-i(\mu_m t + \phi_m)} d_m \op{e}_{m+1,m} \right) 
    e^{-i E_{n'} t} \op{e}_{n',n'} \op{U}_I(t)\\
&=&\sum_m \left(A_m(t) d_m e^{iE_mt} e^{i(\mu_m t+\phi_m)} e^{-i E_{m+1}t}\op{e}_{m,m+1}
          +A_m(t) d_m e^{i E_{m+1} t} e^{-i(\mu_m t +\phi_m)} e^{-i E_m t}\op{e}_{m+1,m}
     \right) \op{U}_I(t)\\
&=&\sum_m A_m(t) d_m 
    \left( e^{i\phi_m}\op{e}_{m,m+1} + e^{-i\phi_m}\op{e}_{m+1,m} \right) \op{U}_I(t)\\
&=&\sum_m A_m(t) d_m \left[ \cos\phi_m \left(\op{e}_{m,m+1}+\op{e}_{m+1,m} \right)
    +i\sin\phi_m \left(\op{e}_{m,m+1}-\op{e}_{m+1,m} \right) \right] \op{U}_I(t)\\
&=&\sum_m A_m(t) d_m 
    \left( -i\cos\phi_m \op{y}_m + i\sin\phi_m \op{x}_m\right) \op{U}_I(t).
\end{eqnarray*}
Hence, multiplying both sides by $-i$ gives
\begin{equation}
  \frac{d\op{U}_I(t)}{dt} = \sum_m A_m(t) d_m 
  \left( \sin\phi_m \op{x}_m - \cos\phi_m \op{y}_m \right) \op{U}_I(t). 
\end{equation}

\section{Lie group decomposition algorithm} 
\label{appendix:Udecomp}
\end{widetext}

To find a decomposition  of the form (\ref{eq:Udecomp}) for a given unitary operator 
$\op{U}$, we set 
\begin{equation}
  \op{U}^{(0)} \equiv e^{-i\Gamma/N} \op{U},
\end{equation}
where $e^{i\Gamma} \equiv \det(\op{U})$, to ensure that $\op{U}^{(0)}\in SU(N)$.  Our
goal is to reduce this matrix $\op{U}^{(0)}$ step by step to a diagonal matrix whose 
diagonal elements are arbitrary phase factors $e^{i\theta_n}$.  Recall that this
reduction is always sufficient if the initial state of the system is an ensemble of 
energy eigenstates.  

Let $U_{ij}^{(0)}$ denote the $i$th row and $j$th column entry in the matrix representation
of $\op{U}^{(0)}$.  In the first step of the decomposition we wish to find a matrix 
\begin{equation}
  \op{W}^{(1)} 
 = \exp\left[-C_1\left(\sin\phi_1\op{x}_1 -\cos\phi_1\op{y}_1\right)\right],
\end{equation}
which is the identity matrix everywhere except for a $2\times 2$ block of the form
\begin{equation}
 \left( \begin{array}{cc} 
       \cos(C_1)               & ie^{i\phi_1} \sin(C_1) \\
       ie^{-i\phi_1} \sin(C_1) & \cos(C_1)   
  \end{array} \right)
\end{equation}
in the top left corner, such that
\begin{equation} \label{eq:W1}
    \op{W}^{(1)} \left(\begin{array}{c} U_{1,N}^{(0)} \\ 
                                        U_{2,N}^{(0)} \\
                                        \vdots
                  \end{array}\right)
  = \left(\begin{array}{c} 0 \\ c \\ \vdots \end{array}\right)
\end{equation}
where $c$ is some complex number.  Noting that 
\[
  U_{1,N}^{(0)} = r_1 e^{i\alpha_1}, \quad
  U_{2,N}^{(0)} = r_2 e^{i\alpha_2}
\]
it can easily be verified that setting 
\begin{equation} \label{eq:Cphi}
   \begin{array}{rcl}
    \phi_k &=& \pi/2+\alpha_1-\alpha_2 \\
    C_k    &=& -\mbox{arccot}(-r_2/r_1)
\end{array}
\end{equation}
achieves (\ref{eq:W1}).  Next we set
\begin{equation}
  \op{U}^{(1)} = \op{W}^{(1)} \op{U}^{(0)}
\end{equation}
and find $\op{W}^{(2)}$ of the form
\begin{equation}
  \op{W}^{(2)} 
 = \exp\left[-C_2\left(\sin\phi_2\op{x}_2 -\cos\phi_2\op{y}_2\right)\right]
\end{equation}
such that
\begin{equation} \label{eq:W2}
    \op{W}^{(2)} \left(\begin{array}{c} 0 \\
                                        U_{2,N}^{(1)} \\ 
                                        U_{3,N}^{(1)} \\
                                        \vdots 
                  \end{array}\right)
  = \left(\begin{array}{c} 0 \\ 0 \\ c \\ \vdots \end{array}\right)
\end{equation}
where $c$ is again some complex number.  Repeating this procedure $N-1$ times leads to
a matrix $\op{U}^{(N-1)}$ whose last column is $(0,\ldots,0,e^{i\theta_N})^T$.  Since
we are not concerned about the phase factor $e^{i\theta_N}$ in this paper, we stop here.
Noting that
\[
 \exp\left[-\frac{\pi}{2}\op{x}_{N-1}\right] \times 
 \exp\left[-\frac{\pi}{2}\left(\sin\phi\op{x}_{N-1}-\cos\phi\op{y}_{N-1}\right)\right]
\]
with $\phi=-\pi/2-\theta_n$ maps $(0,e^{i\theta_{N-1}})^T$ onto $(0,1)^T$, we see that
complete reduction to the identity matrix would require two additional steps at this 
point to eliminate the phase factor $e^{i\theta_N}$, which would result in two additional
control pulses.

Having reduced the last column, we continue with the $(N-1)$st column in the same 
fashion, noting that (since $\op{U}^{(0)}$ is unitary) at most $N-2$ steps will be 
required to reduce the $(N-1)$st column to $(0,\ldots,0,e^{i\theta_{N-1}},0)^T$.
We repeat this procedure until after (at most) $K=N(N-1)/2$ steps $\op{U}^{(0)}$ is
reduced to a diagonal matrix $\mbox{diag}(e^{i\theta_1},\ldots,e^{i\theta_N})$, as
required and we have
\begin{equation}
   \op{W}^{(K)} \cdots \op{W}^{(1)} \op{U}^{(0)} 
 = \mbox{diag}\left(e^{i\theta_1},\ldots,e^{i\theta_N} \right).
\end{equation}
Finally, setting
\begin{equation}
   \op{V}_k \equiv \left(\op{W}^{(K+1-k)}\right)^\dagger
\end{equation}
leads to
\begin{equation}
  \op{U}^{(0)} = \op{V}_K \op{V}_{K-1} \cdots \op{V}_1 
              \mbox{diag}\left(e^{i\theta_1},\ldots,e^{i\theta_N} \right)
\end{equation}
and therefore
\begin{equation}
  \op{U} = \op{V}_K \op{V}_{K-1} \cdots \op{V}_1 \Theta,
\end{equation}
where $\Theta$ is a diagonal matrix
\begin{equation}
  \Theta = e^{i\Gamma/N}\mbox{diag}\left(e^{i\theta_1},\ldots,e^{i\theta_N} \right).
\end{equation}
Note that $\op{U}$ can always be decomposed such that $\Theta$ is the identity matrix.
However, to achieve this goal up to $2(N-1)$ additional terms would be required to 
eliminate the phase factors, which would result in additional control pulses.  While
some applications may indeed require the elimination of these phase factors, the phase
factors are often insignificant and the additional control pulses would be superfluous.
For a more sophisticated decomposition algorithm that requires only very few phases 
the reader is referred to \cite{JCP-preprint}.
\\
\section*{Acknowledgements}
We sincerely thank A.~I.\ Solomon and A.~V.\ Durrant of the Open University for helpful 
discussions and suggestions.  ADG would like to thank the EPSRC for financial support
and VR would like to acknowledge the support of NSF Grant DMS 0072415.  

\bibliography{papers2000,papers9599,papers8089,papers9094}
\end{document}